\documentclass[reprint,amsmath,amssymb,aps,prb,twocolumn,showpacs]{revtex4-1}

\usepackage{color}
\usepackage{graphicx}

\newcommand{\integ}[2]{\int\limits_{#1}^{#2} \!}
\newcommand{\dif}{\, \text{d}}

\begin{document}

\title{Ageing  at the Spin-Glass/Ferromagnet 
Transition: Monte Carlo Simulation  using GPUs}
\author{Markus Manssen}
 \email{markus.manssen@uni-oldenburg.de}
\author{Alexander K. Hartmann}
\affiliation{Institute of Physics, Carl von Ossietzky University,
26111 Oldenburg, Germany}

\date{\today}

\begin{abstract} We study the the non-equilibrium ageing behaviour of the
$\pm J$ Edwards-Anderson model in three dimensions for samples of size 
up to $N=128^3$ and for up to $10^8$ Monte Carlo sweeps. In particular we
are interested in the change of the ageing
when crossing from
the spin-glass phase to the ferromagnetic phase. The necessary long
simulation times are reached by employing a CUDA-based GPU implementation, 
which allows for single-spin flip times as small as
8ps. We
measure typical spin glass correlation functions in space and time to
determine the growing length scale and extract the constituting
exponents. We observe a clear signature of the disorder-driven
equilibrium transition in the non-equilibrium behavior.
\end{abstract}


\maketitle

\section{Introduction} Spin
glasses\cite{binder1986,sgtheorybeyond,fischerspinglasses,nishimori2001,kawashima2013}
are certain magnetic alloys,\cite{mydosh1993} that
possess at low temperatures interesting equilibrium and 
non-equilibrium behaviour which is to a large extend still not well understood.
The low-temperature ordered spin-glass 
phase is characterized by a rough free-energy landscape 
and by slow glassy
dynamics.\cite{bouchaud1998,corberi2011} 
Disorder and frustration in the spin-spin interactions 
were identified as the underlying
principals governing the behavior of spin glasses.
Thus, models mixing positive, i.e., ferromagnetic and negative 
(antiferromagnetic) couplings such as
the mean-field Sherrington-Kirkpatrick
model\cite{sherringtonkirkpatrick} and the Ising-like short-ranged
Edwards-Anderson model\cite{edwardsanderson} were created to
understand the spin-glass behavior. A prevailing topic is still, whether the
replica-symmetry-breaking theory\cite{sgtheorybeyond} arising from the
solution\cite{parisi1979,parisi1983} of the former model can
accurately describe the spin-glass phase of the latter model in three
dimensions. The most prominent competitor is the droplet
theory\cite{fisher1986,fisher1988}, which centers around the eponymous
droplets, low-level excitations from the presumably only two existing
pure states. Numerous publications have dealt with simulations
in\cite{mcmillan1984,kawashima1996,ballesteros2000} as well as out of
equilibrium.\cite{rieger1996,kisker1996,marinari1996,bouchaud1998,komori2000,bouchaud2001,berthier2002,yoshino2002,jaubert2007,belletti2009}

The standard spin-glass models assume an on average equal fraction
of positive and negative couplings. Nevertheless,
when decreasing the fraction of negative bonds in the
Edwards-Anderson model, it exhibits a phase transition at low
temperatures from the aforementioned spin-glass phase to the well
known ferromagnetic phase of the Ising model. This transition has-been 
studied in the typical equilibrium approach to phase
transitions,\cite{nishimori1987} also via ground-state 
calculations.\cite{art_threshold1999}
Nevertheless, concerning the non-equilibrium ``ageing'' behavior,
so far only systems deep in the spin-glass phase have been studied 
extensively, to the knowledge of the authors. Therefore, 
the purpose of this study is to
determine, whether the spin-glass to ferromagnet
 transition is also visible within the
non-equilibrium behaviour. Specifically we will be looking at growing
correlations in space and time and try to explain them in terms of the
dynamical correlation length. The determination and characterization
of this growing length scale in the spin glass phase has been a focus
of many previous
publications\cite{rieger1996,kisker1996,marinari1996,yoshino2002,%
belletti2008,belletti2009},
as there are a few stumbling blocks before dependably measuring it. It
was quickly found, that it appears to follow a power
law\cite{rieger1996,kisker1996} in line with the mean-field theory,
though there is discussion\cite{jaubert2007,belletti2009} whether it
crosses over into the logarithmic growth expected from droplet theory.

Because reaching sufficiently long simulation times is computationally
challenging, even inspiring the adoption of specialized
hardware\cite{janus}, we implemented the model in CUDA\cite{cuda} to
leverage the comparatively high processing power of
GPUs. Quite a few pioneering works have already tested the
feasibility and performance outlook of this platform for the
Ising\cite{preis2009,hawick2009,block2010} and the Edwards-Anderson
model\cite{weigel2012,guidetti2012} but have shown no fruitful
application. Our implementation has been carefully optimized for
tackling the problem at hand efficiently and with limited resources.
This allowed us to study large system sizes of $N=128^3$ up to long
time scales of $10^8$ sweeps.

The remainder of this work is organized as follows:
In section~\ref{sec_model} we describe the used Edwards-Anderson model
and its observables. Section~\ref{sec_implementation} follows with
details on the GPU implementation of the model. The results of the
simulation and their analysis are presented in
section~\ref{sec_results}. We close with our conclusions in
section~\ref{sec_conclusion}.

\section{Model\label{sec_model}}

The Edwards-Anderson model\cite{edwardsanderson} describes a
$D$-dimensional cubic system of side length $L$ containing $N = L^D$
Ising spins $S_i = \pm 1$. Its Hamiltonian is given by
\begin{equation}
 H(S) = -\sum_{\langle i, j \rangle} J_{ij} S_i S_j
\end{equation} where the sum runs over nearest neighbors $\langle i, j \rangle$
and the bonds $J_{ij} = \pm 1$ are drawn from
a bimodal distribution $P(J) = p\delta(J - 1) + (1 - p)\delta(J + 1)$. We use periodic boundary conditions
in all directions. 
The parameter $p\in[0,1]$ controls the fraction of positive and negative bonds.
For $p=1$ the ferromagnet Ising
model is reproduced with a paramagnetic phase at high temperatures and
the well-known ferromagnetic phase at small temperatures $T$ for $D >
1$. On the other hand $p = 0.5$ represents the usual spin glass model
with a low temperature spin-glass phase for $D > 2$. 
We will only be concerned with the case $D = 3$ in the
following, which has the transition temperatures $T_\text{FM} \approx
4.5115$ ($p=1$) \cite{bloete1999} 
and $T_\text{SG} = 1.102(3)$ ($p=0.5$) \cite{baity-jesi2013}, 
respectively.
For intermediate values of $p$ there exists the
phase transition from ferromagnet to spin glass at $p \approx
0.77$ ($T\to 0$)\cite{art_threshold1999}.

Simulations start with random initial configurations emulating a quench 
from infinite temperature. We then examine the system at different 
waiting times $t_w$ (measured in sweeps) after the beginning of the simulation.
The spin glass order parameter is the overlap
\begin{equation}
 q = \frac{1}{N} \sum_i q_i
\end{equation} with $q_i = S_i^{(a)} S_i^{(b)}$ the element-wise
overlap of two replicas $S^{(a)}$ and $S^{(b)}$ with the same bond
configuration $J$, but different initial configurations and thermal histories. In equilibrium, corresponding to 
$t_w\to\infty$, the probability
distribution $P(q)$ is expected to assume a two peak structure below
the transition temperature. In the droplet
theory\cite{fisher1986,fisher1988} this takes the form of two delta
peaks at $\pm q_\text{EA}$, while the mean-field
theory\cite{sgtheorybeyond} has a wider distribution with a plateau of
non-zero probability around $q = 0$. 

To measure the growing length
scale we make use of the spatial four-point correlation
\begin{equation}
 C_4(r, t_w) = \frac{1}{N} \sum_i q_i(t_w) q_{i+r}(t_w)
 \label{eq_C4}
\end{equation}
between two points and two replicas. With ${i+r}$ we denote a spin which
has a spacial distance $r$ from spin $i$.

There exist different approaches to extract a growing coherence 
(or dynamic correlation) length $\xi$
from the four point-correlation function. The first approach is based
on the assumption that $C_4$ follows the functional form\cite{marinari1996}
\begin{equation}
 C_4(r, t_w) \propto r^{-\alpha} g\left(\frac{r}{\xi(t_w)}\right)\,.
 \label{eq_C4form}
\end{equation} The function $g$ is approximately a stretched
exponential $g(x) \sim \exp(-x^\beta)$. 
Extracting $\xi$ works by fitting (\ref {eq_C4form}) to the data of
$C_4$, for various
times $t_w$. The two unknown exponents
($\alpha \approx 0.5$ and $\beta \approx 1.5$ in the spin-glass phase)
complicate the extraction
of $\xi$, which spawned many methods for accomplishing this
task\cite{rieger1996,kisker1996,marinari1996,yoshino2002}. 

As an alternative, notably
Ref.~\onlinecite{belletti2008} introduced the use of integral
estimators for this problem. One uses the integral
\begin{equation}
 I_k(t_w) = \integ{0}{L/2} r^k C_4(r,t_w) \dif r
\end{equation}
to calculate an estimate for the coherence length
\begin{equation}
 \xi_k(t_w) = \frac{I_{k+1}(t_w)}{I_k(t_w)} \sim \xi(t_w).
 \label{eq_xik}
\end{equation}
The choice of $k$ determines which regions of the 
function $C_4$ contribute most to the estimate. 
Ref.~\onlinecite{belletti2009} recommends $k = 1$ to get the best 
trade-off between systematic errors for small values of $k$ and larger 
influence of statistical error for higher values. 

Another 
observable of interest we use to study the aging behavior
around the ferromagnet-spin glass transition is the autocorrelation
\begin{equation}
 C(t, t_w) = \frac{1}{N} \sum_i S_i(t_w) S_i(t_w + t)
 \label{eq_C}
\end{equation} between two points in time separated by a time
difference $t$ in reference to the waiting time $t_w$. It is assumed
to split into two parts. The first is a quasi-equilibrated
part for small $t \ll t_w$, that takes the
form\cite{rieger1996,kisker1996,komori2000,berthier2002,jaubert2007}
of a power law 
\begin{equation}
C_\text{eq}(t) \propto t^{-x}\,,
\label{eq:power:x}
\end{equation} 
with another
characteristic exponent $x$.
 For longer times $t \gg t_w$
the ageing part $C_\text{age}(t,t_w) = f\left( \xi(t_w + t) / \xi(t_w)
\right)$ can trivially be expected\cite{bouchaud1998,corberi2011} to
depend only on the ratio of the coherence lengths at the two
different times. An additive decomposition $C(t,t_w) = C_\text{eq}(t)
+ C_\text{age}(t,t_w)$ is favored by theoretical
arguments\cite{berthier2002,jaubert2007,corberi2011}. But we will make
use of a multiplicative decomposition $C(t,t_w) = C_\text{eq}(t) \cdot
C_\text{age}(t,t_w)$, as this seems to work better, even though it is
only expected to hold in the critical
region.\cite{berthier2002,corberi2011}
\section{Implementation\label{sec_implementation}}

 We implemented a Metropolis Monte Carlo
simulation\cite{newman1999} of the model for Nvidia GPUs using the
CUDA C programming interface\cite{cuda}, as was first detailed in
Ref.~\onlinecite{manssenmaster}. For explanation of the
GPU-related terms used in the following we refer to the CUDA
Programming Guide\cite{cuda} or a textbook like
Ref.~\onlinecite{kirk2012}.  In order to perform the update of a spin
$S_i$ one has to calculate the flipping probability
\begin{equation}
 p_\text{accept} = \min \left(1, \exp\left(-\frac{2}{T}
\sum_{j \in N(i)} J_{ij} S_i S_j\right)\right)
 \label{eq_metropolis}
\end{equation} incorporating the coupling of the spin $i$ to the local
field generated by its direct neighbours $N(i)$ on the lattice. 
Since GPUs are designed
to keep their large number of simple processors busy with many,
preferably independent processes at once, a sequential implementation of a
single-spin-flip algorithm is ill-suited for GPUs. So in order to attain a
parallel algorithm suitable for this architecture we adopted a
standard checkerboard update scheme. In a first step all ``white fields'' of
the system are updated followed by the other half of the system in a
second step. Both add up to a single sweep of the system corresponding
to one time step. Each half-step is performed in its own CUDA  kernel call
to ensure global synchronization.

\begin{figure}
 \includegraphics[width=0.45\textwidth]{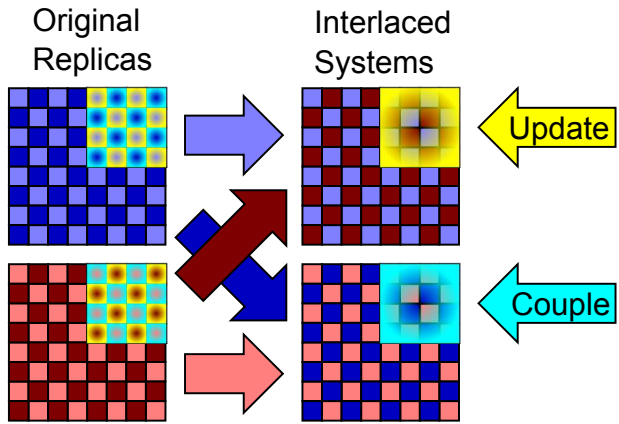}
 \caption{(Color online) \label{fig:layout} Memory layout in the GPU:
Two independent replicas (left), having the same bonds, 
are simulated in a checker board manner: During each
iteration either all ``black'' or all ``white'' sites are updated. Thus,
within the GPU memory, one memory area (top right) 
holds all updated sites, e.g., the ``white''
sites of replica one and the ``black'' sites of replica two, while 
another memory range (lower right) contains all neighboring 
interaction partners. 
After one half-sweep is completed, the role of updated sites and  neighboring sites 
are swapped.}
\end{figure}

A straight-forward implementation of this update scheme however would
be inefficient as the GPU memory is optimized for reading large bulks
of neighboring memory cells at once (``memory coalescing''). To
circumvent this one could relocate the spins to two different memory
regions as was done in Ref.~\onlinecite{hawick2009}. For a
simpler method we instead simulate the two replicas we need for
calculating (\ref{eq_C4}) simultaneously and swap the ``black spins''
between the two lattices to get what we will call the ``interlaced
checkerboard layout'', see Fig.\ \ref{fig:layout}.
 This way all spins in one lattice can be
updated at the same time, while all neighbours, they are coupled to,
reside in the other lattice. Basically the same approach was also used
in Ref.~\onlinecite{janus} just by virtue of the simplifications it
introduces. Specifically the spins' indices remain unchanged and we
can use the same bonds for both update steps. Since the bonds $J_{ij}$
are symmetric, we only need to store the left/up bonds of a spin and
they can be read efficiently via texture memory. The joint neighbors
of the updated spins are loaded into shared memory so they can be
shared between the members of a thread block to calculate the flip
probability (\ref{eq_metropolis}). We choose dimensions of
$32\times4\times2$ for the thread blocks in the GPU thread hierarchy.

We also employed 64bit-multispin-coding meaning, that the spins taking
values $\pm 1$ are coded as single bits and 64 of them are stored
together in a 64bit-variable. The same applies for the corresponding
bonds. We choose spins from the same position in 64 different samples,
which is sometimes known as asynchronous multispin-coding. Bit
operations are used to perform the update for all bit-coded spins at
the same time. We only need to differentiate between a few possible
cases of spin alignments using boolean logic at the bit level. Then we
look up the precalculated flipping probabilities from constant memory
and apply them for the matching bits. It is customary to save
computational effort by using only one random number for multiple
samples.  As no suitable and efficient random number generators were
available at the time of implementation, we established a 1024bit
variant of a Xorshift generator\cite{manssen2012}. The generator was
optimized for generating a single random number per thread and kernel,
as was needed here. With this complete implementation we reach
single-spin-flip times of $\approx 8\text{ps}$ on a GeForce GTX 570
GPU. Of the prior
implementations\cite{hawick2009,preis2009,weigel2010,block2010,weigel2011,weigel2012,guidetti2012}
of Ising and Edwards-Anderson model only the one by
Weigel\cite{weigel2012} is faster. But it uses multi-hit updates,
which means each thread block updates for several flip-attempts
per spin, thus requiring the costly global synchronization less frequently. This
is no problem if one is interested in equilibrium properties, but
 this changes the dynamics, as, e.g.,  visible by the reduced growth of
correlations. Therefore, that is undesirable when one wants to actually study
the non-equilibrium dynamic ageing properties, as in the current work.

For the current work we only had access to two GPUs and consequently
designed this approach for maximum efficiency per GPU. However if one
had access to a larger number, it would be preferable to be able to
distribute samples better among GPUs. For this purpose one can simply
reduce the number of samples to $M = 2^m$ ($m<6$) by storing $64 / M$ spins
from different positions per sample in a multispin. A simple way is to
split the system into $64 / M$ equal parts along one dimension and
assign spins at the same relative positions to the bits $\{i, i + M, i
+ 2 M, \ldots\}$ in the same multispin. This effectively makes it look
and work like a smaller system with the caveat, that when coupling
over the ``periodic boundary'' one has to rotate the bits of the
multispin variable by $M$ positions. The computational overhead for
this change is negligible but it has two other problems. Firstly this
effectively shrinks our systems which might result in low occupancy
and efficiency of the GPUs processors. But the bigger problem is, that
we cannot use the same random number for spins from the same
sample. Thus we have to generate $64 / M$ random numbers per kernel
instead of just one. Because of this requirement, the 
synchronous multispin-coding
corresponding to $M = 1$ is very inefficient, and more balanced
choice like $M = 8$ is preferable.
\section{Results\label{sec_results}}

\begin{figure}
 \includegraphics[width=0.5\textwidth]{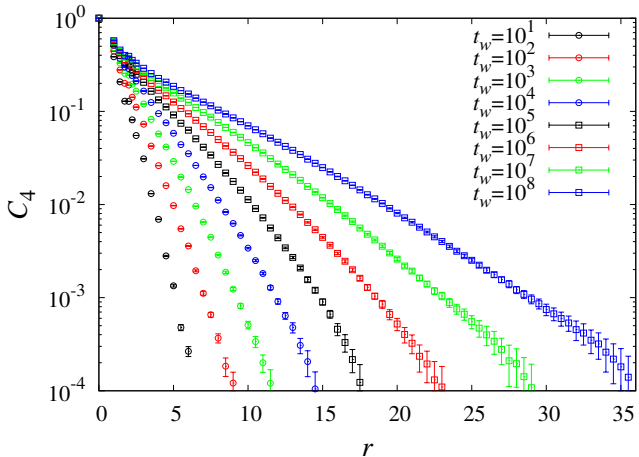}
 \caption{(Color online) Spatial correlation $C_4$ over the distance $r$ at 
different waiting times $t_w$ for a $128^3$ system at $p = 0.5$. 
Multiple close points were merged to give a clearer picture.\label{fig_C4}}
\end{figure}

\begin{figure}
 \includegraphics[width=0.5\textwidth]{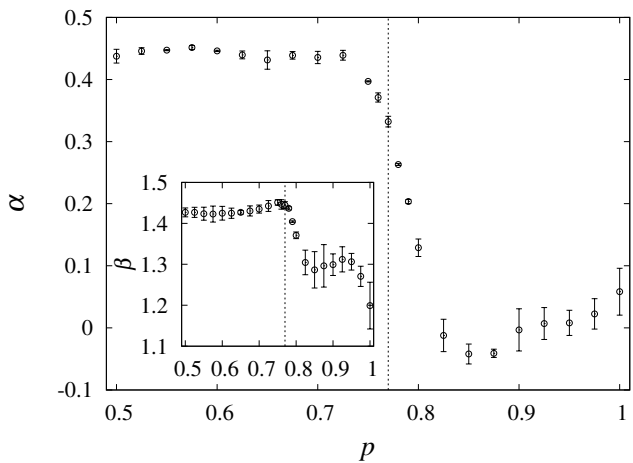}
 \caption{Scaling exponent $\alpha$ of the spatial correlation over the bond probability $p$ for a $128^3$ system. The vertical line marks $p = p_c$. 
Inset: Associated scaling exponent $\beta$.\label{fig_alphabeta}}
\end{figure}

We simulated a total of 192 samples of randomly
initialized $128^3$ systems with two replicas each. The simulations
were performed on two GeForce GTX 570. We took the parameters $T =
0.8$ and $p \in [0.5, 1]$ and performed $10^8$ time steps, which takes
about 63h per batch of 64 multispin-coded samples. At the measure
points the whole system configurations where simply stored to hard
disk and the generated data was later post-processed to gain access to
the correlation functions.  

An exemplary spatial correlation $C_4$
from (\ref{eq_C4}) for $p = 0.5$ is shown in Fig.~\ref{fig_C4}.
The two replicas utilized for the calculation make
correlations visible despite the model's inherent disorder. The curves
show a seemingly exponential decay for larger distances. The steeper
gradient for small distances is incorporated in the scaling form
(\ref{eq_C4form}) with the power law $r^{-\alpha}$. As one would
expect, the correlations spread to larger distances as time passes
suggesting a growing length scale.

  Our first approach to extracting
this coherence length $\xi$ is a fit of (\ref{eq_C4form}) with the
stretched exponential form for $g$. However, the values $\alpha \approx 0.5$ 
and $\beta \approx 1.5$ which are suitable deep in 
the spin-glass phase ($p\approx 0.5$)
might depend on the value of $p$. To get the most consistent
values at a particular $p$ we performed multifits of the curves for
all different $t_w \ge 10$ at once, i.e. with universal values 
of $\alpha$ and $\beta$ (independent of $t_w$),
but individual values $\xi(t_w)$. Naturally the choice of points included in the
fit can have an influence on the outcome. As such we generally
restricted it to $r \ge 3$ and specifically found the cleanest
results, when only using points at integer-valued distances $r$. Those are
always located along the lattice axes. But as a reference we did the
same fits also with all $r \ge 3$ and use these for calculating our
errorbars for $\alpha$, $\beta$ and $\xi$. In detail we estimate our
errors as the difference between the fit results for our restricted
point set and the larger one plus both of the normal statistical
errors from the two different fits. Still, this might underestimate
the errors a bit, because for multifits statistical independence of the
data is assumed, while in our case the measurements are from the same
runs, just at different waiting times $t_w$, thus they are correlated.

The results for  both exponents $\alpha$ and $\beta$ as a
function of the probability $p$ are shown in
Fig.~\ref{fig_alphabeta}.  A strong change can be seen around the phase
transition $p_c \approx 0.77$ from $0.4 < \alpha < 0.5$ and $1.4 <
\beta < 1.45$ in the spin glass phase to $\alpha \approx 0$ and $\beta
\approx 1.3$ in the ferromagnetic phase. Thus, the equilibrium phase
disorder driven phase transition is well visible in the analysis
of the non-equilibrium ageing behavior.
Note that 
when  getting closer to $p = 1$, i.e., in the ferromagnetic phase,
the system quickly develops long-range order. Nevertheless, due to the low 
temperature, this is not an equilibrium magnetized configuration but a system
with two large domains separated by a long-living domain wall.
Thus, on the one hand, when addressing the range where the coherence
length is small,  we only have the first few time $t_w$ available 
to work with, i.e., the fits according to (\ref {eq_C4form})
generally cannot be fitted as well at later times.

\begin{figure}
 \includegraphics[width=0.5\textwidth]{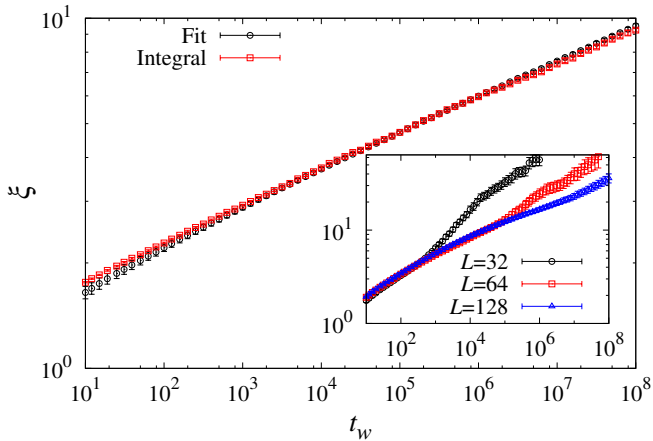}
 \caption{(Color online) Coherence length $\xi$ as a function of
 the waiting times $t_w$ 
for a $128^3$ system at $p = 0.5$. The values are calculated by 
fitting and integral estimation respectively. {Inset: 
Results by fitting estimation for different side lengths 
$L \in \{32,64,128\}$ at $p = 0.9$.}\label{fig_xi}}
\end{figure}

\begin{figure}
 \includegraphics[width=0.5\textwidth]{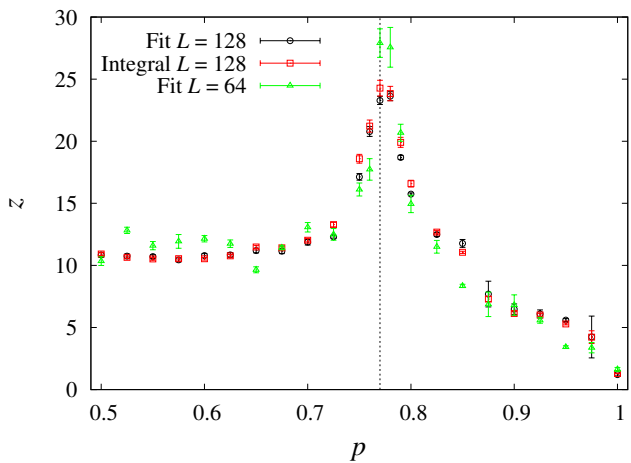}
 \caption{(Color online) Power law exponent $z$ of the coherence
length as a function of the bond probability $p$ for $L=128^3$ and $L=64$. 
For $L=128$,
the two 
curves correspond to the fitting and integral estimation of $\xi$ 
respectively, which agree pretty well.
To give an impression of the finite-size effects, also the result 
from the fitting approach for $L=64$
(exhibiting more systematic errors due to the small system size) is included.
 The vertical line marks $p = p_c$.\label{fig_z}}
\end{figure}

The second
approach\cite{belletti2008,belletti2009} for calculating
$\xi$ uses the integral estimation $\xi_1$ according to
(\ref{eq_xik}). Like in the original work we take the integrals up to
the point, where the value of $C_4$ first becomes smaller than three
times its error, and approximate the rest of the integrals with our
fitted function. Fig.~\ref{fig_xi} shows 
results for the coherence length $\xi$ as a function of the
waiting time $t_w$ for both different approaches, respectively.  
As is visible from the figure, both methods agree
well for a large stretch of waiting times but disagree close to the
beginning and the end. While the integral results look closer to a
power law, the fit results give a bit higher estimates for $\xi$ at
the end and bend down at short times.   A grave
problem arises with finite size effects in the ferromagnetic phase, as
can be seen in the inset of Fig.~\ref{fig_xi}. When $\xi$ becomes
comparable to the system's own length scale $L$, the values get
overestimated and the systems start to actually equilibrate. This
means we would need to go to even larger systems to get better results
in the ferromagnetic phase at these temperatures. 

Anyway, to study the
dependence on the concentration $p$ of ferromagnetic bonds, we fitted
power laws of the form $\xi(t_w)\sim t_w^{1/z}$, which is the most-simple
yet widespread approach. 
 This power-law behavior however is subdued at the beginning. So for
fitting purposes we found, it is best to add a constant term, that
then usually takes negative values. The determined exponents $z$ for
different $p$ are shown in Fig.~\ref{fig_z} for both methods of
extracting $\xi$.  The phase transition can again be seen. Starting
from a constant value between 10 and 11 in the spin glass phase $z$
has a peak around the phase transition before decreasing in the
ferromagnetic phase. 
Larger values of $z$ correspond to slower growth of correlations
and consequently overall slower behaviour and equilibration. So it
fits expectations that the spin glass phase has much higher values of $z$ than
the ferromagnetic phase. But interestingly we can see an even more
pronounced slowdown in the critical region around the 
disorder-driven phase transition.
Here the dynamics are so slow that one could expect even just a
logarithmic growth of the coherence length with waiting time. Thus, we
also tried right at $p\approx p_c$ a fit to the functional form 
$\xi(t_w)\sim \log(t_w/t_0)^{\tilde z}$ with parameters $t_0$ and $\tilde z$. 
The fit worked as well as the
power-law fit, thus, we are not able to distinguish a very slow
power-law growth from a logarithmic growth here. For values of $p$ 
close to one, note
 that the finite-size effect in $\xi$ affects the
values of $z$, which causes a deviation from
the expected $z = 2$ for $p = 1.0$. 
Furthermore, we tried to extrapolate the value of $p_c$ from, this
data. For this purpose, we
fitted a Gaussian to the peaks of the $L=64$ and $L=128$ data, while the data
for $L=32$ has a very pronounced peak at $p=0.78$ (spacing $\Delta p = 0.01$).
As a result, we obtain estimates $p_c(L=32)=0.78(1)$, $p_c(L=64) = 0.777(2)$
and $p_c(L=128)=0.772(1)$. Thus, no pronounced finite-size effect is visible,
see also the peak of the $L=64$ date in Fig.\ \ref{fig_z}.
This can be expected, since we analyzed non-equilibrium data in the time
interval where the coherence length is much smaller than the system size.
Thus, it does not make much sense to try to extrapolate $p_c$ for
large systems sizes, which appears anyway not necessary since the 
obtained values for $p_c$ are very well compatible with the finite-size 
estimate from equilibrium studies.

\begin{figure}
 \includegraphics[width=0.5\textwidth]{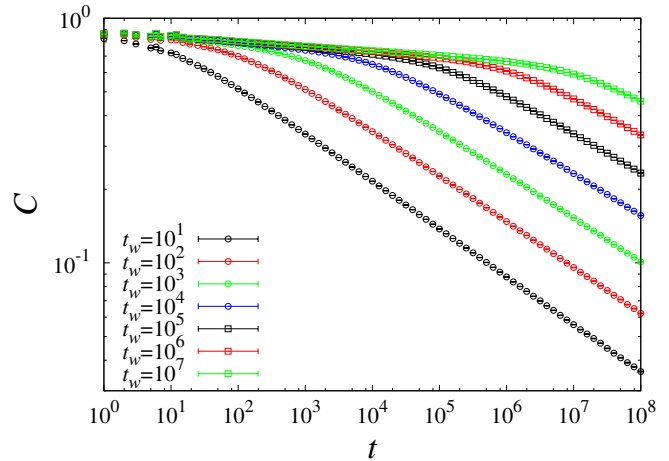}
 \caption{(Color online) Autocorrelation $C$ as a function of 
the time distance $t$ at different waiting times $t_w$ for a $128^3$ 
system at $p = 0.5$.\label{fig_C}}
\end{figure}

\begin{figure}
 \includegraphics[width=0.5\textwidth]{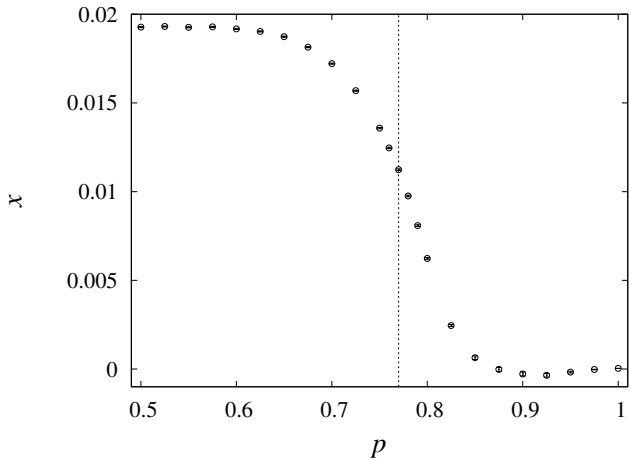}
 \caption{Equilibrium exponent $x$ of the autocorrelation as a function of the 
bond probability $p$ for a $128^3$ system. The vertical line 
marks $p = p_c$.\label{fig_x}}
\end{figure}

In order to validate these results for $\xi$ we will take a look at
the autocorrelation $C$ from (\ref{eq_C}). An example for $p = 0.5$ is
shown in Fig.~\ref{fig_C}.  Transitions can be seen around $t = t_w$,
respectively,
from the equilibrium regime with slow algebraic decay to the ageing
regime with faster decay. This corresponds to the notion that up to
time $t_w$ the system is equilibrated on length scales of size $\xi(t_w)$
and it takes time $t>t_w$ to make a spin feel that a system is not equilibrated
at longer length scales.

To obtain the so-called equilibrium exponent
$x$ defined in (\ref{eq:power:x}) we fit corresponding power laws 
to regions of different width,
the smallest being $t \in[100, t_w/1000]$. As the values agree well
for the different fitting regions and $t_w$, we take the mean as our
result. We show in Fig.~\ref{fig_x} the equilibrium exponent $x$ at
different probabilities.  The exponent drops at the phase transition
from a finite value $x \approx 0.019$ for the spin glass $p = 0.5$ to
basically zero for the ferromagnet $p = 1.0$. This simply
means that the equilibrium correlation in the ferromagnetic phase at
low temperatures is almost constant compared to the spin glass
phase and we do not expect to see as much drift. 

\begin{figure}
 \includegraphics[width=0.5\textwidth]{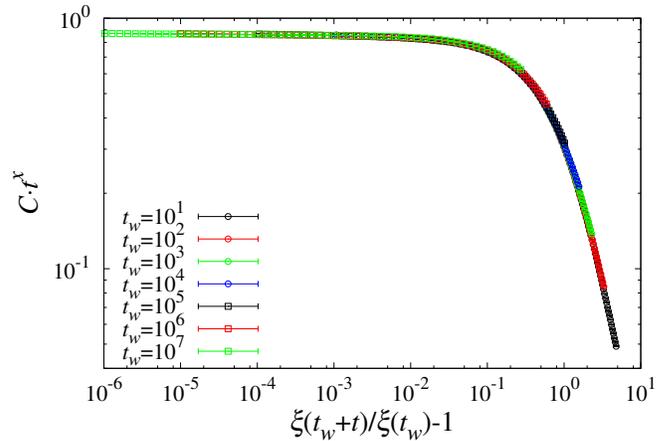}
 \caption{(Color online) Collapse of the autocorrelation $C$ at 
different waiting times $t_w$ for a $128^3$ system at $p = 0.5$. 
The values for $\xi$ are interpolated from the fitting results. 
$x$ is adjusted for best collapse.\label{fig_C_col}}
\end{figure} 

 Next we test the
assumption, that the ageing part scales with $\xi(t_w+t)/\xi(t_w)$ by
trying a collapse in Fig.~\ref{fig_C_col} using a multiplicative
decomposition $C(t_w, t) = C_\text{eq}(t) \cdot C_\text{age}(t_w, t)$.
We used the results for $\xi$  previously obtained using the
fit approach. This gave a better
collapse than the integral results especially for small
$t_w$. For the display in  Fig.~\ref{fig_C_col}, we subtract $1$ from our 
abscissa  to make the collapse of
values for $t \ll t_w$ and thus $\xi(t_w + t) \approx \xi(t_w)$ better
visible. As visible from the figure, the quality of the collapse is very good.
The optimal value of $x \approx 0.016$ for the collapse is a
bit smaller than the fit result shown in Fig.~\ref{fig_x}
 and generally seems to be 
susceptible to the particular form of $\xi$. Note that we tried also
collapses with
an additive decomposition $C(t_w, t) = C_\text{eq}(t) +
C_\text{age}(t_w, t)$ but this resulted for all cases in a worse
overlap of the curves even with its additional free parameter.  


\section{Conclusion}
\label{sec_conclusion} 

We were able to perform relatively
long simulations of the Edwards-Anderson model at low temperatures for
multiple different probabilities from the spin glass phase up to the
ferromagnetic one. This was made possible largely through the use of
general-purpose GPU (CUDA) computing, which has become feasible in recent
years. Thus, the CUDA-based approach allows for very fast simulations 
of spin glasses at much cheaper costs
compared to standard CPU systems or even compared to specific FPGA-based 
 hardware like the Janus computer.\cite{janus}

The ageing behavior of the spin glass phase seen in previous
study was reproduced well. The main purpose of our work was to study
the ageing behavior of the system as a function of the variable
fraction of ferromagnetic bounds. 
We could easily detect the transition from the spin-glass
phase to the ferromagnet, when
altering the bond probability $p$. This was visible in all quantities
we measured. Note that only
at the extreme end of the
ferromagnetic phase finite size effects began to complicate matters
considerably and we cannot give exact results there. Our entry point,
the spatial correlation, contains information about the growing length
scale of a system but, because we lack an explicit form for it,
getting reliable values is difficult. The integral estimators taken
from Ref.~\onlinecite{belletti2008,belletti2009} have proven useful
but do not seem to work as well at short times. A multifit of the
assumed form (\ref{eq_C4form}) can help out for these cases. It also
turns out the matching form changes noticeably when crossing the phase
transition line. While the exponent $\beta$ of the stretched
exponential only lowers slightly, the other exponent $\alpha$
basically vanishes in the ferromagnetic phase, thus arriving at a
simpler form.

The autocorrelation exhibits a quasi-equilibrated part with power law
behavior in the spin-glass phase. The equilibrium exponent $x$ also
vanishes in the ferromagnetic phase, which does not exhibit as much
rearrangement in equilibrium. The ageing part on the other hand can be
scaled well with the quotient $\xi(t_w+t)/\xi(t_w)$, also giving
credence to their calculated values. The assumption of power law
growth works for the coherence length albeit with a small correction
for early values. However we did not delve into testing a crossover to
logarithmic growth. The power law exponent $z$ is naturally higher for
the slower dynamics of the spin glass. But it also shows additional
slowdown in the critical region around the phase transition. In
essence all findings agree with the general expectations
that in the ferromagnet the
equilibrium state is more uniform and stationary and systems can
arrive there much faster. Nevertheless, we were quite surprised how well
the equilibrium disorder-driven 
transition shows up when measuring the non-equilibrium ageing properties.

However our whole approach was focused on the long time simulation of
spin glasses and as such we could not get as good results for the
ferromagnet. Because of the faster evolution a different emphasis
would have to be put to fare better. Also it can be seen as a bit
questionable to make use of the trick of recycling random numbers for
different samples without a strong influence of the disorder. 
In any case, beyond the physical results, 
 the developed implementation and
analysis methods can be used to proceed further efficiently 
with investigations of the equilibrium and non-equilibrium behavior of the 
random bond model.

\section{Acknowledgments}

We thank Martin Weigel and 
A. Peter Young for many interesting discussions and helpful
suggestions.

\bibliography{bib_art_general}

\end{document}